\documentstyle[12pt,openbib,psfig]{article}
\newcommand{\f}{\frac}

\newcommand{\p}{\partial}
\newcommand{\La}{\Lambda}

\newcommand{\be}{\beta}
\newcommand{\al}{\alpha}

\newcommand{\va}{\varepsilon}
\def\be{\begin{equation}}
\def\ee{\end{equation}}

\begin{document}

\title{Incompressibility of strange matter $^*$.}

\author{ Monika Sinha $^{1,2,3}$, Manjari Bagchi $^{1,2}$, Jishnu Dey
$^{4,5**}$ Mira Dey $^{1,5**}$, \\Subharthi Ray $^6$ and
Siddhartha Bhowmick $^7$}

\date{}
\maketitle

\begin{abstract}
{Strange stars (ReSS) calculated from a realistic equation of
state (EOS), that incorporate chiral symmetry restoration as well
as deconfinement at high density\cite{d98} show compact objects
in the mass radius curve. We compare our calculations of
incompressibility for this EOS with that of nuclear matter. One of
the nuclear matter EOS has a continuous transition to ud-matter at
about five times normal density. Another nuclear matter EOS
incorporates density dependent coupling constants. From a look at
the consequent velocity of sound, it is found that the transition
to ud-matter seems necessary. }
\end{abstract}

\vskip 1cm

 Keywords{ compact stars~--~realistic strange
stars~--~dense matter~--~elementary particles ~--~ equation of
state}

$^1$ Dept. of Physics, Presidency College, 86/1 College Street,
Kolkata
700073,~India; $^2$ monika2003@vsnl.net, mnj2003@vsnl.net

$^3$ CSIR NET Fellow;

$^4$ UGC Research Professor, Dept. of Physics, Maulana Azad
College, 8 Rafi Ahmed Kidwai Road, Kolkata 700013,~India;

 $^5$ Associate, IUCAA, Pune, India;

$^{**}$ deyjm@giascl01.vsnl.net.in

$^6$ Inter University Centre for Astronomy and Astrophysics, Post
bag 4, Ganeshkhind, Pune 411007, India; sray@iucaa.ernet.in

$^7${Department of Physics, Barasat Govt. College, Kolkata 700
124, India.}

{$^*$ Work supported  in  part  by DST  grant no. SP/S2/K-03/01,
Govt. of India.}.

\section{Introduction}

An exciting issue of modern astrophysics is the possible
existence of a family of compact stars made entirely of
deconfined u,d,s quark matter or ``strange matter" (SM) and
thereby denominated strange stars (SS). They differ from neutron
stars, where quarks are confined within neutrons, protons and
eventually within other hadrons (hadronic matter stars). The
possible existence of SS is a direct consequence of the so called
strange matter hypothesis\cite{bodmer}, according to which the
energy per baryon of SM would be less than the lowest energy per
baryon found in nuclei, which is about 930 MeV for $Fe^{56}$.
Also, the ordinary state of matter, in which quarks are confined
within hadrons, is a metastable state. Of course, the hypothesis
does not conflict with the existence of atomic nuclei as
conglomerates of nucleons, or with the stability of ordinary
matter\cite{farhi,madsen,Bombaci}.

The best observational evidence for the existence of quark stars
come from some compact objects, the X-Ray burst sources
SAX~J1808.4$-$3658 (the SAX in short) and 4U~1728$-$34, the X-ray
pulsar Her X-1 and the superburster 4U~1820$-$30. The first is
the most stable pulsating X-ray source known to man as of now.
This star is claimed to have a mass $M* \sim 1.4 ~M_\odot~ $ and
a radius of about 7 kms \cite{Li99a}. Coupled to this claim are
the various other evidences for the existence of SS, such as the
possible explanation of the two kHz quasi-periodic oscillations
in 4U 1728~-~34 \cite{Li99b} and the quark-nova explanation for
$\gamma$ ray bursts \cite{boma}.

\begin{figure}[htbp]
\centerline{\psfig{figure=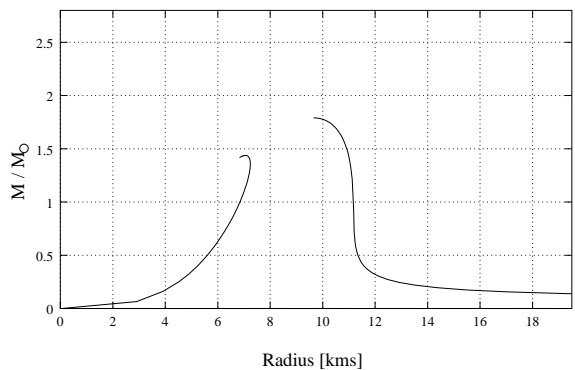,height=5cm}}\vskip .5cm
\caption{\label{MR} Mass and radius of stable stars with the strange
star EOS (left curve) and neutron star EOS (right curve), which
are solutions of the Tolman-Oppenheimer-Volkoff (TOV) equations
of general relativity. Note that while the self sustained strange
star systems can have small masses and radii, the neutron stars
have larger radii for smaller masses since they are bound by
gravitation alone.}
\end{figure}

The expected behaviour of SS is directly opposite to that of a
neutron star as Fig.(\ref{MR}) shows. The mass of 4U~1728$-$34 is
claimed to be less than  1.1 $M_\odot$ in Li et al. \cite{Li99b},
which places it much lower in the M-R plot and thus it could be
still gaining mass and is not expected to be as stable as the SAX.
So for example, there is a clear answer \cite{singhini} to the
question posed by Franco \cite{fr01}: why are the pulsations of
SAX not attenuated, as they are in 4U~1728$-$34 ?

From a basic point of view the equation of state for SM should be
calculated solving QCD at finite density. As we know, such a
fundamental approach is presently not feasible even if one takes
recourse to the large colour philosophy of 't Hooft \cite{'t
Hooft}. A way out was found by  Witten \cite{wit} when he
suggested that one can borrow a phenomenological potential from
the meson sector and use it for baryonic matter. Therefore, one
has to rely on phenomenological models. In this work, we use
different equations of state (EOS) of SM proposed by Dey et al
\cite{d98} using the phenomenological Richardson potential. Other
variants are now being proposed, for example the chromo dielectric
model calculations of Malheiro et al.\cite{mal}.

\begin{figure}[htbp]
\centerline{\psfig{figure=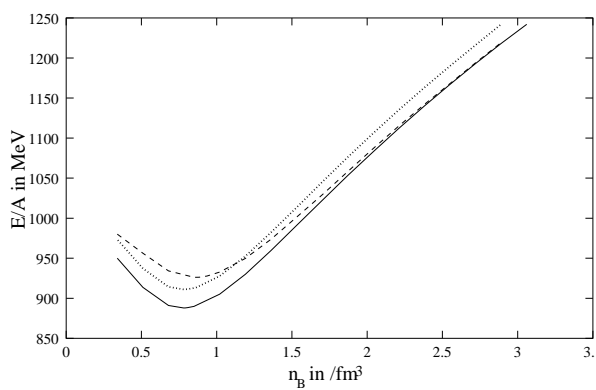,height=5cm}}
\vskip .5cm
\caption{Strange matter EOS employed by D98 show respective
stable points. The solid line for is EOS1, the dotted line for
EOS2 and the dashed line for EOS3. All have the minimum at energy
per baryon less than that of $Fe^{56}$} \label{efit}
\end{figure}

Fig.(\ref{efit}) shows the energy per baryon for the EOS of
\cite{d98}. One of them (eos1, SS1 of \cite{Li99a}) has a minimum
at E/A = 888.8 $MeV$ compared to 930.4 of $Fe^{56}$, i.e., as
much as 40 $MeV$ below. The other two have this minimum at 911
$MeV$ and 926 $MeV$, respectively, both less than the normal
density of nuclear matter. The pressure at this point is zero and
this marks the surface of the star in the implementation of the
well known TOV equation. These curves clearly show that the
system can fluctuate about this minimum, so that the zero
pressure point can vary.

\section{Incompressibility : its implication for Witten's Cosmic
Separation of Phase scenario.}

In nuclear physics incompressibility is defined as\cite{bdp} \be
K~=~ 9 \f{\p} {\p n}\left(n^2\f{\p\varepsilon}{\p n}\right)~,
\label{com}\ee where $\va~=~E/A$ is the energy per particle of
the nuclear matter and $n$ is the number density. The relation of
K with bulk modulus B is \be K=\f{9B}n~~.\label{bulk}\ee $K$ has
been calculated in many models. In particular, Bhaduri et al
\cite{bdp} used the non - relativistic constituent quark model,
as well as the bag model, to calculate $K_A$ as a function of $\
n$ for the nucleon and the delta. They found that the nucleon has
an incompressibility $ K_N$ of about 1200 MeV, about six times
that of nuclear matter. They also suggested that at high density
$K_A$ matches onto quark gas incompressibility.

The velocity of sound in units of light velocity c is given by
\be v = \sqrt{K/9 \va}\label{vel}.\ee

The simple models of quark matter considered in \cite{d98} use a
Hamiltonian with an interquark potential with two parts, a scalar
component (the density dependent mass term) and a vector
potential originating from gluon exchanges. In the absence of an
exact evaluation from QCD, this vector part is borrowed from meson
phenomenology \cite{richardson}. In common with the
phenomenological bag model, it has built in asymptotic freedom
and quark confinement (linear). In order to restore the
approximate chiral symmetry of QCD at high densities, an {\it
ansatz} is used for the constituent masses, viz.,
\begin{equation}
M(n) = M_Q~sech\left(\nu\f{n}{n_0}\right), \label{sech}
\end{equation}
where $n_0$ is normal nuclear matter density and $\nu$ is a
parameter. There may be several EOS's for different choices of
parameters employed to obtain the EOS. Some of them are given in
the table (\ref{eos}). There $\al_s$ is perturbative quark gluon
coupling and $\La$ is the scale parameter appeared in the vector
potential. Changing other parameters one can obtain more EOS's.
The table (\ref{eos}) also shows the masses($M_G$),radii (R) and
the number density($n_s$) at star surface of maximum mass strange
stars obtained from the corresponding EOS's. The surface of the
stars occur at the minimum $\va$ where pressure is zero.

\begin{table}[htbp]
\caption{\label{eos}Parameters for the three EOS}
\vskip .5cm
\begin{center}
\begin{tabular}{lccccccccr}
EOS&$\nu$&$\al_s$&$\La$&$M_Q$&$M_G/M_{\odot}$&R& $n_s/n_o$\\
&&&&MeV&&(km)&\\ \hline
EOS1&0.333&0.20&100&310&1.437&7.055&4.586\\ \hline
EOS2&0.333&0.25&100&310&1.410&6.95&4.595\\ \hline
EOS3&0.286&0.20&100&310&1.325&6.5187&5.048\\
\end{tabular}
\end{center}
\end{table}

\begin{figure}[htbp]
\vskip .5cm \centerline{\psfig{figure=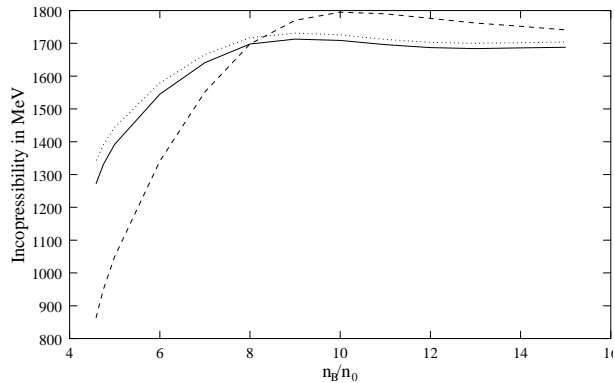,height=5cm}}
\vskip .5cm \caption{\label{inceos} Incompressibility as a
function of density ratio for EOS's for different EOS's given in
table(\ref{eos}). The solid curve for EOS1, the dotted curve for
EOS2 and the dashed curve for EOS3.}
\end{figure}

The general behaviour of the curves is relatively insensitive to
the parameter $\nu$ in $M(n)$ as well as the gluon mass, as
evident from the figure(\ref{inceos}).

\begin{figure}[htbp]
\vskip .5cm
\centerline{\psfig{figure=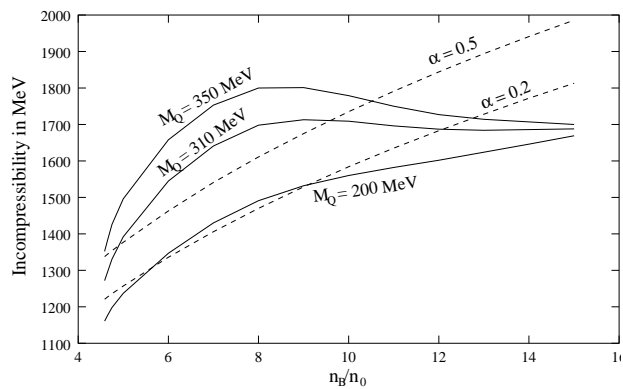,height=5cm}}
\vskip .5cm
\caption{\label{incmas} Incompressibility as a function of density ratio for
EOS's with different constituent mass as parameter.
Dashed lines correspond to
perturbative massless three flavour quark gas with different values of
$\alpha_s$ (see \cite{bdp,baym}).}
\end{figure}

In figure (\ref{incmas}), $K$ with three values of $M_Q$ implying
different running masses, $M(n)$, is plotted as a function of the
density expressed by its ratio to $n_0$. Given for comparison, is
the incompressibility $K_q$ of a perturbative massless three
flavour quark gas consisting of zero mass current quarks
\cite{bdp} using the energy expression given in \cite{baym} to
order $\alpha_s ^2$.

 It can be seen that as $M_Q$ decreases,
the nature of the relation approaches the perturbative case of
\cite{bdp}. At high density our incompressibility and that due to
Baym \cite{baym} matches, showing the onset of chiral symmetry
restoration. In EOS1 for uds matter, the minimum of $\varepsilon$
occurs at about $4.586 ~ n_0$. nucleation may occur at a density
less than this value of $n$. This corresponds to a radius of about
$0.67~fm$ for a baryon. For EOS1 we find $K$ to be $1.293~GeV$ per
quark at the star surface.

It is encouraging to see that this roughly matches with the
compressibility $K_N$ so that no `phase expands explosively'. In
the Cosmic Separation of Phase scenario, Witten \cite{wit} had
indicated at the outset that he had assumed the process of phase
transition to occur smoothly without important departure from
equilibrium. If the two phases were compressed with significantly
different rates, there would be inhomogenieties set up.

But near the star surface at $n~\sim~4$ to $5~n_0$ the matter is
more incompressible showing a stiffer surface. This is in keeping
with the stability of strange stars observed analytically with
the Vaidya-Tikekar metric by Sharma, Mukherjee, Dey and Dey
\cite{smdd}.

\begin{figure}[htbp]
\vskip .5cm
\centerline{\psfig{figure=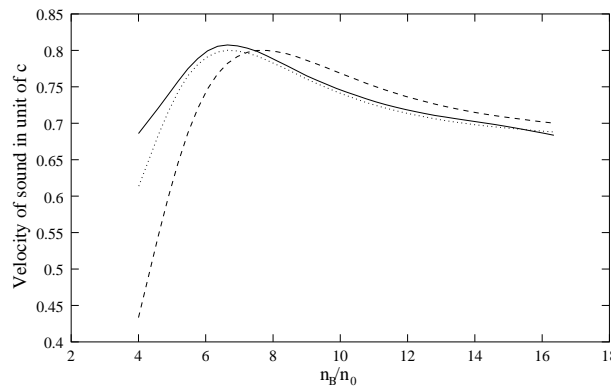,height=5cm}}
\vskip .5cm
\caption{Velocity of sound, $v_s$ as a function of density
ratio. The solid line is for EOS1, the dotted for EOS2 and the
dashed for EOS3.} \label{vels}
\end{figure}

The velocity of sound, $v_s$ peaks somewhere around the middle of
the star and then falls off. We show it for three different EOS in
fig.(\ref{vels}) with parameters given in Table (\ref{eos}).

Next we turn to the model of Zimanyi and Moskowski \cite{zm},
where the coupling constants are density dependent. It was shown
\cite{dddm} that the quark condensate derived from this model via
the Hellmann-Feynman theorem is physically acceptable in this
model. Details may be found in \cite{dddm} but the essence is
that in the more conventional Walecka model the condensate
increases with density whereas in QCD a decrease is expected.

\begin{figure}[htbp]
\vskip .5cm \centerline{\psfig{figure=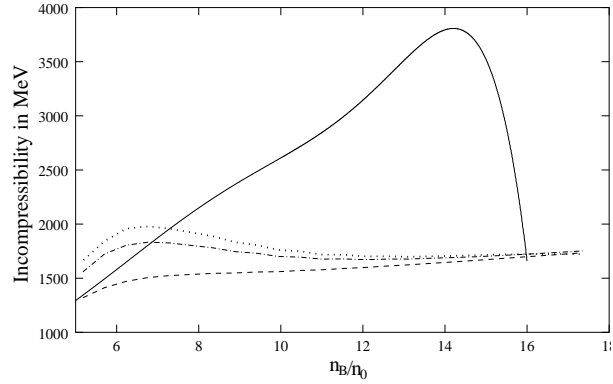,height=5cm}} \vskip
.5cm \caption{Incompressibility of neutron matter(ZM model) with
three strange matter EOS's for different constituent mass as
parameter as a function of density ratio. The solid line is for
neutron matter the dotted for $M_Q=350 MeV$, the dash-dot for
$M_Q=310 MeV$ and the dashed for $M_Q=200 MeV$.} \label{sns}
\end{figure}

\begin{figure}[htbp]
\vskip .5cm \centerline{\psfig{figure=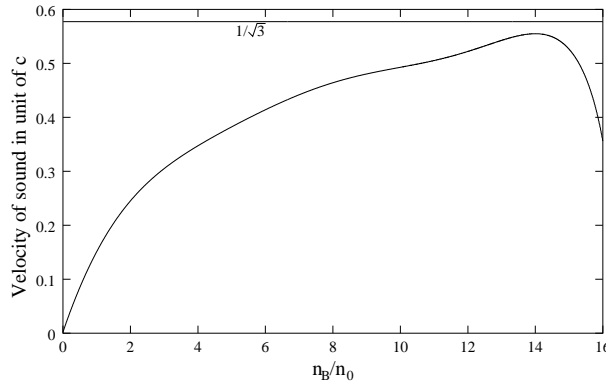,height=5cm}} \vskip
.5cm \caption{Velocity of sound, $v_s$ as a function of density
ratio for ZM neutron matter.} \label{velnu}
\end{figure}

In a recent paper \cite{krein} Krein and Vizcarra (KV in short)
have put forward an EOS for nuclear matter which exhibits a
transition from hadronic to quark matter. KV start from a
microscopic quark-meson coupling Hamiltonian with a density
dependent quark-quark interaction and construct an effective
quark-hadron Hamiltonian which contains interactions that lead to
quark deconfinement at sufficiently high densities. At low
densities, their model is equivalent to a nuclear matter with
confined quarks, i.e., a system of non-overlapping baryons
interacting through effective scalar and vector meson degrees of
freedom, while at very high densities it is ud quark matter. The
$K_{NM}$ at the saturation density is fitted to be $248~MeV$. This
EOS also gives a smooth phase transition of quark into nuclear
matter and thus, conforms to Witten's assumption. Interestingly
enough, the transition takes place at about $\sim 5 n_0$.

The KV model does not incorporate strange quarks so that
comparison with our EOS is not directly meaningful. However it is
quite possible that the signal results of the KV calculations
mean that quark degrees of freedom lower the energy already at
the ud level and once the possibility of strange quarks is
considered the binding exceeds that of $Fe^{56}$. An extension of
KV with strange quarks is in progress\footnote{G. Krein, by
e-mail}.

\begin{figure}[htbp]
\centerline{\psfig{figure=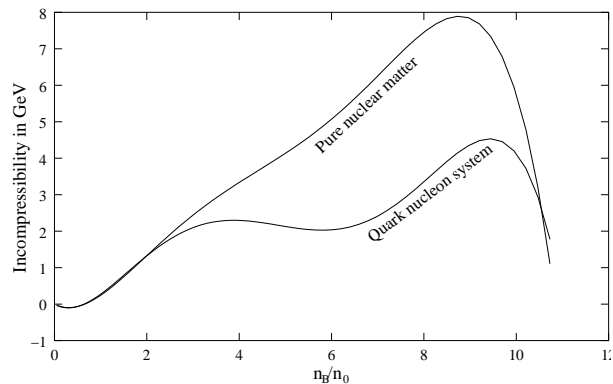,height=5cm}}
\vskip .5cm
\caption{Incompressibility as a function of density ratio for
pure nuclear matter and quark-nucleon system.} \label{krin}
\end{figure}

Results obtained from the KV calculation are presented
simultaneously. The incompressibility shows softening and the
velocity of sound decreases when quark degrees of freedom open
up, as expected. At $\sim~5n_0$ $K_{qN}$ is about $2~GeV$ in
qualitative agreement with our value.

\begin{figure}[htbp]
\centerline{\psfig{figure=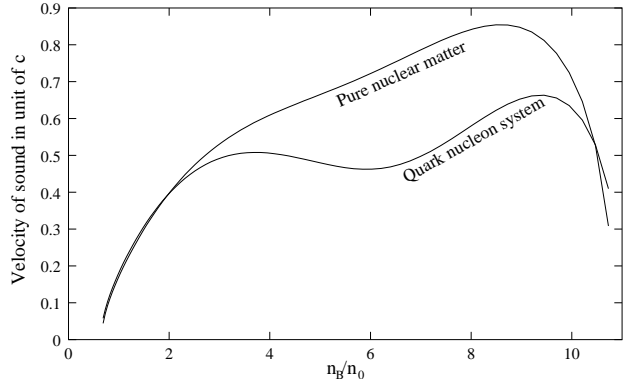,height=5cm}}
\vskip .5cm
\caption{The velocity of sound in pure nuclear matter and in
quark-nuclear system as a function of density ratio. For pure
nuclear matter at $9 n_0$ the sound velocity is too close
to that of light c, whereas for quark-nuclear system it is much less,
about 0.5 c.}
\label{krvs1}
\end{figure}

Note that in our EOS we also have strange quarks reducing the
value of $K$. Fig (\ref{krvs1}) shows the $v_s$ as a function of
$n_B/n_0$ for both kinds of EOS. The EOS with quarks shows a
lowering of $v_s$.

\section{Discussions.}

~~~In our model the density is large even at the surface
($4.586~n_0$ for $EOS1$) so that the quarks in the fermi sea have
high momentum on the average. Therefore, they often come close
together. It was shown by Bailin and Love that in certain colour
channels \cite{bl}, the short range one gluon exchange spin-spin
interaction can produce extra attraction, forming diquarks. In
s-state the pair can be formed only in flavour antisymmetric
state. The flavour antisymmetric diquark may form in colour
symmetric(6) and spin symmetric (triplet) state, or in colour
antisymmetric ($\bar3$) and spin  antisymmetric (singlet) state.
The potential strength of the latter is 6 times the former. The
N-$\Delta$ mass splitting is obtainable from the one gluon
exchange potential, the delta function part of which is smeared to
a Gaussian,

\be H_{i,j}~=~-
80.51~\sigma^3(\lambda_i.\lambda_j)(S_i.S_j)~e^{-\sigma^2
r_{ij}^2}. \label{diq} \ee

In the above, the strength of this spin dependent part is taken
from Dey and Dey(1984) \cite{dd}.  The pairing energy of ud pairs
in the spin singlet and colour $\bar3$ state is $-3.84~MeV$
\cite{MNRAS} \footnote{With this pairing energy superbursts
observed from some astrophysical x-ray sources has been explained
very well.}. Our model does not predict diquarks that are
permanent - since the quarks must have high momentum transfer if
they are to interact strongly with a force that is short-range.
The formation and breaking of pairs  give rise to endothermic and
exothermic processes leading to fast cooling and superbursts
respectively.

We also note that the lowering of energy is not very large, as
compared  to the approximately 42 $MeV$ energy difference per
baryon (930.6 $Fe^{56}$ to $888~MeV$), seen in EOS1. So we  do not
expect any drastic change in the incompressibility or the velocity
of sound due to diquark pairing, nor do we expect a phase
transition to a colour-flavour locked state.

We are grateful to the referee for inviting a comment on quark
pairing in the strange matter scenario.

\end{document}